\documentclass[proceedings, preprint]{rmaa}
\usepackage{paralist}
\usepackage{psfrag,color}

\SetYear{2008}
\SetConfTitle{IV Reuni\'on\,�sobre Astronom\'\i{}a Din\'amica en Latino Am\'erica }
\title{Dynamical friction force exerted on spherical bodies}
\author{
  Esquivel, O.,\altaffilmark{1,2} 
  Fuchs, B.\altaffilmark{1}}
\altaffiltext{1}{Astronomisches Rechen-Institut am Zentrum f\"ur Astronomie der 
Universit\"at Heidelberg,\\
M\"onchhofstra{\ss}e 12-14, 69120 Heidelberg, Germany}
\altaffiltext{2}{Fellow of International Max-Planck Research School for
Astronomy and Cosmic Physics, Heidelberg (esquivel@ari.uni-heidelberg.de).}

\shortauthor{Esquivel \& Fuchs}
\shorttitle{Dynamical friction on spherical bodies}

\listofauthors{O. Esquivel, \& B. Fuchs}
\indexauthor{Esquivel, O.}
\indexauthor{Fuchs, B.}
\abstract{Following a wave-mechanical treatment we calculate the drag force
exerted by an infinite homogeneous background of stars on a perturber
as this makes its way through the system. We recover Chandrasekhar's 
classical dynamical friction (DF) law with a modified Coulomb logarithm.
We take into account a range of models that encompasses all plausible density distributions
for satellite galaxies by considering the DF exerted on a Plummer sphere and 
a perturber having a Hernquist profile. It is shown 
that the shape of the perturber affects only the exact form of the Coulomb
logarithm. The latter converges on small scales, because encounters of the test
and field stars with impact parameters less than the size of the massive 
perturber become inefficient. We confirm this way earlier results based on the
impulse approximation of small angle scatterings.} 
\resumen{Siguiendo un enfoque mec\'anico-ondular calculamos la fuerza de arrastre ejercida 
por un sistema homog\'eneo e infinito de estrellas de fondo sobre pertubador 
mientras este se mueve a trav\'es del sistema.
Recuperamos la f\'ormula cl\'asica para la fuerza de fricci\'on (FF) derivada por
Chandrasekhar, pero con un logaritmo de Coulomb modificado. 
Al estimar la FF ejercida sobre una esfera de Plummer y un perturbador que posee un perfil tipo
Hernquist, consideramos un intervalo de modelos que abarca toda distribuci\'on plausible de sat\'elites
gal\'acticos. Se muestra que la configuraci\'on del perturbador afecta \'unicamente la forma exacta del
logaritmo de Coulomb. Tal logaritmo converge a peque\~{n}as escalas porque los encuentros
entre la part\'\i{}cula de prueba y las estrellas de fondo cuyos par\'ametros de impacto son inferiores
al tama\~{n}o del perturbador masivo resultan ineficientes. Comprobamos as\'\i{} los 
resultados previos basados en la aproximaci\'on de impulso de peque\~{n}as deflecciones angulares.}

\addkeyword{methods: analytical}
\addkeyword{galaxies: kinematics and dynamics}
\begin{document}
\maketitle

\section{Introduction}

The process of dynamical friction (DF) is one of the most classical
and fundamental problems encountered in the description of the evolution
of almost all astrophysical systems. From the critical momentum exchange
in a protoplanet-protoplanetary disk set up, passing through the problem
of satellites in galaxies to galaxies in large clusters, proper understanding
of DF is a prerequisite to more ambitious attempts at constructing physically
justified models. 

In his seminal paper \citet{1943ApJ....97..255C}
envisaged the scenario of a sequence of consecutive gravitational two--body
encounters of test and field stars in order to calculate the drag force
(cf.~\citealt{H73} for a modern presentation). 
In particular, application of DF to calculate the rate of a sinking satellite has received special
attention, and efforts have been made to include more general background distributions.
However, as for the perturber itself, \citet{1976MNRAS.174...19W} has been the only one to 
consider a more realistic finite-size perturber to compute analytically the DF based on
an impulse-approximation approach.
His main result was a modification of the Coulomb logarithm so that 
it does not diverge anymore at small scales, because gravitational encounters
at impact parameters smaller than the size of the perturbing body become ineffective.
Here we rigorously calculate the drag force exerted on different bodies following
the approach of both \citet{1968SvA....11..873M} and \citet{K72} who determined the 
"polarization cloud" created in the background medium as a massive object was making its
way through the system. 
This method can be simply understood as linear and angular momentum exchange in
stellar systems (\citealt{1972MNRAS.157....1L}, \citealt{1976PhR....24..315D}, \citealt{1984MNRAS.209..729T},
\citealt{2004A&A...419..941F}), and has been extensively used in plasma physics (cf. \citealt{ST92}).
It is shown in section 4 below that the shape of the perturber affects only 
the exact form of the Coulomb logarithm. As concrete examples we calculate the
drag force exerted on a Plummer sphere and on a sphere with the density
distribution of a \citet{1990ApJ...356..359H} profile, respectively, and compare this with the
drag force exerted on a point mass.
\section{A wave-mechanical treatment}
We assume an infinite homogenous distribution of field stars on isotropic 
straight--line orbits. The response of the system of background stars to the
perturbation due to a massive perturber is determined by solving the
linearized Boltzmann equation
\begin{equation}
\frac{\partial f_1}{\partial t} + \sum_{i=1}^3 {\rm v}_{\rm i} 
\frac{\partial f_1}{\partial {\rm x}_{\rm i}}  - 
\frac{\partial \Phi_1}{\partial {\rm x}_{\rm i}}
\frac{\partial f_0}{\partial {\rm v}_{\rm i}} = 0\,
\label{eq1}
\end{equation}
where $\Phi_1$ denotes the gravitational potential of the perturber and 
$f_1$ and $f_0$ are the perturbed and unperturbed distribution function of the field 
stars in phase space. By Fourier-transforming both the perturbations $f_1$ and the potential $\Phi_1$
(whose form are $f_{\omega,{\bf k}};\, \Phi_{\omega,{\bf k}} \, \mathrm{exp}i[\omega t + 
{\bf k}\cdot{\bf x}]\,$; where $\omega$ and ${\bf k}$ denote the frequency and wave vector of the 
Fourier components) the solution of Boltzmann equation is greatly facilitated.
Without loss of generality the spatial coordinates 
${\rm x}_{\rm i}$ and the corresponding velocity components ${\rm v}_{\rm i}$
can be oriented with one axis parallel to the direction of the wave vector.
The Boltzmann equation (\ref{eq1}) takes then the form
\begin{equation}
\omega f_{\omega,{\bf k}} + \upsilon kf_{\omega,{\bf k}} - 
k\Phi_{\omega,{\bf k}}\frac{\partial f_0}{\partial \upsilon} = 0
\label{eq3}
\end{equation}
with $k=|{\bf k}|$ and $\upsilon$ denoting the velocity component parallel
to ${\bf k}$. Equation (\ref{eq3}) has been integrated over the two velocity
components perpendicular to ${\bf k}$. In the following we assume for the
field stars always a Gaussian velocity distribution function, to find the solution
\begin{equation}
f_{\omega,{\bf k}}=-\frac{kv}{\omega+k\upsilon}
\frac{n_{\rm b}}{\sqrt{2\pi}\sigma^3}e^{-\frac{\upsilon^2}{2\sigma^2}} 
\Phi_{\omega,{\bf k}}\,.
\label{eq5}
\end{equation}
where $n_{\rm b}$ denotes the spatial density of the field stars.
Integrating eq.~(\ref{eq5}) over the $\upsilon$--velocity leads to the 
density distribution of the induced polarization cloud. This has been calculated
here without taking into account the self--gravity of the background medium. 
However, \citet{2004A&A...419..941F} has shown that in linear approximation the effects of 
self--gravity are not important for the dynamics of the polarization cloud.
\section{Potentials of the perturbing bodies}
To start with, we consider in our analysis the potential of a
point mass, which moves with the velocity ${\rm v}_0$ along the y--axis,
\begin{equation}
\Phi_1=-\frac{Gm}{\sqrt{{\rm x}^2 + ({\rm y}-{\rm v}_0 t)^2 + {\rm z}^2}}\,.
\label{eq6}
\end{equation}
Its Fourier-Transform can be calculated using formulae (3.754) and (6.561) of
\citet{GR00} as
\begin{equation}
\Phi_{\bf k}=-\frac{Gm}{2\pi^2}\frac{1}{k^2}e^{-i{\rm k}_{\rm y}{\rm v}_0 t}\,.
\label{eq7}
\end{equation}
Next, the potential (\ref{eq6}) is generalized to
\begin{equation}
\Phi_1=-\frac{Gm}{\sqrt{r_0^2 + {\rm x}^2 + ({\rm y}-{\rm v}_0 t)^2 + 
{\rm z}^2}}\,,
\label{eq8}
\end{equation}
which corresponds to an extended body with the mass distribution of a Plummer
sphere ($\rho \propto 1/r_0^3 (1+r^2/r_0^2)^{-\frac{5}{2}}\,$;\ \citealt{1987gady.book.....B})
The Fourier transform of a moving Plummer sphere is given by
\begin{equation}
\Phi_{\bf k}=-\frac{Gm}{2\pi^2}\frac{r_0}{k}K_1(kr_0)
e^{-i{\rm k}_{\rm y}{\rm v}_0 t}\,,
\label{eq9}
\end{equation}
where $K_1$ denotes the modified Bessel function of the second kind.
As third example we consider a perturber which has the mass density 
distribution of a Hernquist profile ($\rho \propto (r_0/r)(r_0+r)^{-3}$;\ \citet{1990ApJ...356..359H}).
Its gravitational potential is given by
\begin{equation}
\Phi_1 = - \frac{Gm}{r_0+r}\,.
\label{eq10}
\end{equation}
The Fourier transform of a moving Hernquist sphere can be calculated using
eq.~(3.722) of \citet{GR00}\footnote{We use the identity 
$\mathrm{sin}kr=-\frac{1}{r}\frac{\partial}{\partial k}\mathrm{cos}kr$ in eq.~(3.722).}
leading to
\begin{eqnarray}
\Phi_{\bf k}=-\frac{Gm}{2\pi^2}\frac{1}{k^2}
[1 + kr_0\,\mathrm{cos}(kr_0)\mathrm{si}(kr_0)\nonumber \\
 - kr_0\,\mathrm{sin}(kr_0)\mathrm{ci}(kr_0)] e^{-i{\rm k}_{\rm y}{\rm v}_0 t}\,,
\label{eq11}
\end{eqnarray}
where si and ci denote the sine- and cosine-integrals, respectively.
The model of a Plummer sphere has often been used in numerical simulations of 
the accretion and their eventual disruption of satellite galaxies in massive 
parent galaxies. Plummer spheres have constant density cores, whereas 
numerical simulations of the formation of galactic haloes in cold dark 
matter cosmology show that dark haloes may have a central density cusp 
(\citealt{1997ApJ...490..493N}). Thus models of a Plummer or a Hernquist sphere should 
encompass the range of plausible models for satellite galaxies. The density
in both models falls off radially steeper than found in the cold dark matter 
galaxy cosmogony simulations. This mimics the tidal truncation of satellite 
galaxies in the gravitational field of their parent galaxies. 
\section{Dynamical friction}
The ensemble of stars is accelerated by the moving perturber as
\begin{equation}
<\dot{\bf v}> = - \int d^3{\rm x} \int d^3{\rm v} f({\bf x},{\bf v}) \nabla
\Phi_1 \,,
\label{eq11a}
\end{equation}
where $f$ denotes the full distribution function $f=f_0+f_1$. The 
contribution from $f_0$ cancels out, and introducing the Fourier transforms we find
\begin{eqnarray}
<\dot{\bf v}>=
-\int d^3{\rm x} \int d^3{\rm v} 
\int d^3{\rm k}\,  i {\bf k}\Phi_{\bf k}
\nonumber \\ \times  e^{i[\omega t +{\bf k}\cdot{\bf x})]} 
\int d^3{\rm k'} f_{\bf k'} 
e^{i[\omega' t + {\bf k}'\cdot{\bf x}]} \,.
\label{eq12}
\end{eqnarray}
From symmetry reasons the  acceleration vector $<\dot{\bf v}>$ is 
expected to be oriented along the y--axis.
In eq.~(\ref{eq12}) the frequency $\omega$, and
similarly $\omega'$, is given according to eqns.~(\ref{eq7}),(\ref{eq9}) and 
(\ref{eq11}) by $\omega=-{\rm k}_{\rm y} {\rm v}_0-i\lambda$ where we have
introduced following Landau's rule a negative imaginary part, which we will let 
go to zero in the following.
Moreover, $\Phi^*_{\bf k} = \Phi_{-{\bf k}}$ so that the potential is a
real quantity. Equation (\ref{eq12}) simplifies to
\begin{eqnarray}
<\dot{\bf v}>=(2\pi)^3\int d^3{\rm v}
\int d^3{\rm k}\, i{\bf k}
\Phi_{-{\bf k}} f_{\bf k} e^{2\lambda t}\,.
\label{eq13}
\end{eqnarray}
Using expression (\ref{eq5}) and taking the limit $\lambda\to 0$ we get 
\begin{eqnarray}
<\dot{\bf v}>=\frac{(2\pi)^{5/2} \pi n_{\rm b}}{\sigma^3}
\int d^3{\rm k}\,|\Phi_{\bf k}|^2\frac{\bf k}{k}
\, {\rm k}_{\rm y}{\rm v}_0 
e^{-\frac{({\rm k}_{\rm y}{\rm v}_0)^2}{2k^2\sigma^2}}\,.
\label{eq16}
\end{eqnarray}
The Fourier transform of any potential with spherical symmetry depends only 
on $k=|\mathbf{k}|$. Thus it follows immediately from eq. (\ref{eq16}) that 
indeed the two acceleration components ($<\dot{\rm v}_{\rm x}>\ =\ <\dot{\rm v}_{\rm z}>\ =\ 0$)
as anticipated.
Only in the direction of motion of the perturber there is a net effect. 
According to Newton's third law the drag force exerted on the perturber is 
given by $m\dot{\rm v}=-m_{\rm b}<\dot{\rm v}_{\rm y}>$ where $m_{\rm b}$ is 
the mass of a background particle, so that the drag force
is anti--parallel to the velocity of the perturber. In order to evaluate the
integrals over the wave numbers in eq.~(\ref{eq16}) it is advantageous to switch
from Cartesian form ${\rm k}_{\rm x}, {\rm k}_{\rm y}, {\rm k}_{\rm z}$ to
a mixed representation ${\rm k}_{\rm y}$, 
$k = \sqrt{{\rm k}_{\rm x}^2+{\rm k}_{\rm y}^2+{\rm k}_{\rm z}^2}$,
$\mathrm{arctan}({\rm k}_{\rm x}/{\rm k}_{\rm z})$, and we obtain for the
deceleration the general result
\begin{eqnarray}
\dot{\rm v}=-\frac{4\pi G^2 m m_{\rm b} n_{\rm b}}{{\rm v}_0^2}
\left[ \mathrm{erf}
\left( \frac{{\rm v}_0}{\sqrt{2}\sigma}\right)-\sqrt{\frac{2}{\pi}}
\frac{{\rm v}_0}{\sigma}e^{-\frac{{\rm v}_0^2}{2\sigma^2}}\right]\mathrm{ln}\Lambda
\label{eq18}
\end{eqnarray}
where erf denotes the usual error function. The Coulomb logarithm is defined as
\begin{equation}
\mathrm{ln}\Lambda = \frac{4\pi^4}{G^2m^2}\int^{k_{\rm max}}_{k_{\rm min}}
dk\ k^3|\Phi_{\rm k}|^2
\label{eq19}
\end{equation}
In the case of a point mass formula (\ref{eq7}) implies 
$\Lambda=k_{\rm max}/k_{\rm min}$.
This result was first obtained in this form by \citet{K72} and is identical 
to Chandrasekhar's formula (\citealt{1943ApJ....97..255C}), if $m+m_b\approx m$. The Coulomb
logarithm diverges in the familiar way both on small and large scales, i.e.~at
$k_{\rm max}^{-1}$ and $k_{\rm min}^{-1}$, respectively.
The Coulomb logarithm of the dynamical friction force exerted on a Plummer
sphere can be calculated by inserting eq.~(\ref{eq9}) into (\ref{eq19}) 
leading to
\begin{equation}
\mathrm{ln}\Lambda = r_0^2\int^{\infty}_{k_{\rm min}}dk\ k\, K^2_1(r_0k)\,.
\label{eq20}
\end{equation}
If the Plummer radius $r_0$ shrinks to zero, expression (\ref{eq20}) changes
smoothly into the Coulomb logarithm of a point mass, because 
$\lim_{r_0\to 0}\ r_0K_1(k_0k)=k^{-1}$. The integral over the square of the
Bessel functions in eq.~(\ref{eq20}) can be evaluated using formula (5.54) of 
\citet{GR00},
\begin{equation}
\mathrm{ln}\Lambda = -\frac{ r_0^2k_{\rm min}^2}{2} \left[ K^2_1(r_0k_{\rm min})
-K_0(r_0k_{\rm min})K_2(r_0k_{\rm min})\right]
\label{eq21}
\end{equation}
which is approximately
\begin{equation} 
 \mathrm{ln}\Lambda  \approx  -1/2-\mathrm{ln}(r_0k_{\rm min})\,,
\label{eq21a}
\end{equation}
in the limit of $ r_0k_{\rm min}\ll 1$.
This modified Coulomb logarithm converges on small scales precisely as 
found by \citet{1976MNRAS.174...19W}, but still diverges on large scales. A natural cut--off 
will be then the size of the stellar system under consideration.
For a perturber with the density distribution of a Hernquist profile we find
a Coulomb logarithm of the form
\begin{eqnarray}
\mathrm{ln}\Lambda=\int^{\infty}_{k_{\rm min}}dk\ \frac{1}{k}[1 + r_0k\,
\mathrm{cos}(r_0k)\mathrm{si}(r_0k) \nonumber\\
- r_0k\,\mathrm{sin}(r_0k)\mathrm{ci}(r_0k)]^2\,.
\label{eq22}
\end{eqnarray}
It can be shown using the asymptotic expansions of the sine-- and 
cosine--integrals given by \citet{1972hmf..book.....A} that the integrand in expression 
(\ref{eq22}) falls off at large $k$ as $4r_0(r_0k)^{-5}$. Thus the Coulomb 
logarithm converges at small scales. This is expected because, although the 
density distribution has an inner density cusp, the deflecting mass `seen' 
by a field star with a small
impact parameter scales with square of the impact parameter. At small wave 
numbers a Taylor expansion shows that the square bracket in expression
(\ref{eq22}) approaches 1 so that we find a logarithmic divergence of the 
Coulomb logarithm as in the case of the Plummer sphere.
\begin{figure}[!t]\centering
\vspace{0pt}
\includegraphics[width=1.0\columnwidth]{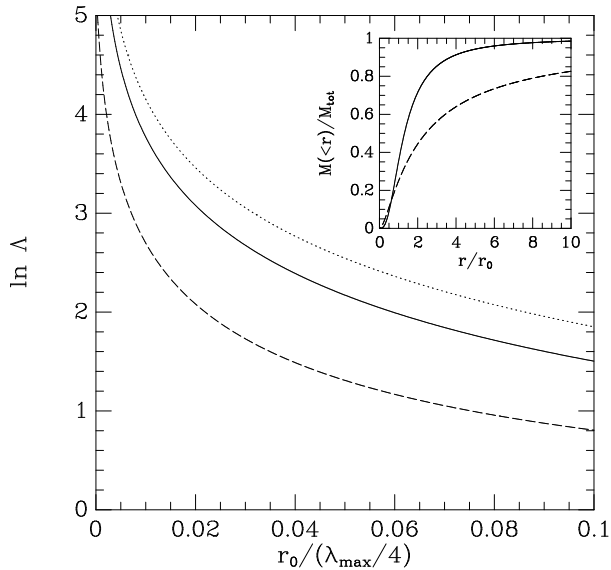}
\caption{Coulomb logarithms of the Plummer sphere (solid line), and a sphere 
with a Hernquist density profile (dashed line). The dotted line indicates
$-{\rm ln}(2 \pi r_0/\lambda_{\rm max})$. $r_0$ denotes the radial scale lengths
of the spheres and $\lambda_{\rm max}$ is the upper cut--off of the wavelength 
of the density perturbations (see text). The inlet shows the cumulative mass 
distributions of the Plummer and Hernquist models.}
\label{fig1}
\end{figure}
In Fig.~1 we illustrate the Coulomb logarithms of the Plummer sphere and a
sphere with a Hernquist density distribution according to eqns.~(\ref{eq21}) 
and (\ref{eq22}) as function of $r_0k_{\rm min}$.
Fig.~1 shows clearly that at given mass small sized perturbers experience 
a stronger dynamical friction force than larger ones. Since the cut-off of the 
wave number at $k_{\rm min}$ is determined by the radial extent of the stellar
system, which corresponds roughly to one half of the largest subtended wave 
length $\lambda_{\rm max}=2\pi /k_{\rm min}$, we use actually 
$r_0/(\lambda_{\rm max}/4)$ as abscissa in Fig.~1. For comparison we have also 
drawn $- ln(r_0k_{\rm min})$ in Fig.~1. The insert shows the cumulative mass 
distributions of both mass models. The half mass radius of the 
Hernquist model measured in units of $r_0$ is about twice as that of the 
Plummer sphere.
Thus for a proper comparison of the drag forces exerted on a Plummer sphere 
and a sphere with a Hernquist density profile the dashed line in Fig.~1 should 
be stretched by a factor of about 2 towards the right. But it is clear from 
Fig.~1 that the drag force exerted on a Plummer is always larger that the drag 
on a sphere with a Hernquist density profile. This is to be expected because of 
its shallower density profile. The logarithm $- ln(r_0k_{\rm min})$, although 
being the asymptotic expansion of the Coulomb logarithms (\ref{eq21}) and 
(\ref{eq22}) for $r_0k_{\rm min} \rightarrow 0$, is not a good approximation at 
larger  $r_0k_{\rm min}$. There is a systematic off--set relative to 
the true Coulomb logarithms which is given explicitely in eq.~(\ref{eq21a}) for 
the case of the Plummer sphere.
It could be argued that perturbers can be deformed by tidal fields. 
However, its effect is expected to be small (a higher order effect). 
There is a further effect if the perturbers are gravitationally bound systems
themselves like globular clusters or dwarf satellite galaxies. 
Such objects can and do loose mass due to tidal shocking and both effects can even be of comparable magnitude.
Finally, we want to mention that our analysis can be extended, in a straightforward way, to
anisotropic velocity distributions of the field stars. \citet{2005AA...444..455F} have shown that
the velocity dispersion in the solution of the Boltzmann 
equation (\ref{eq5}) is replaced by an effective velocity dispersion which depends on the semi-axes of the velocity 
ellipsoid and its orientation relative to the wave vector ${\bf k}$. 
We intend to present our results in an forthcoming paper (Esquivel \& Fuchs, in preparation).
\section{Acknowledgements}
 O.E.~gratefully acknowledges financial support by the International-Max-Planck-Research-School for Astronomy
and Cosmic Physics at the University of Heidelberg. ]


\begin{thebibliography}
\bibitem[Abramowitz \& Stegun(1972)]{1972hmf..book.....A}
    Abramowitz M., Stegun I.A.\ 1972, Handbook of Mathematical Functions,
    Dover, New York
\bibitem[Binney \& Tremaine(1987)]{1987gady.book.....B}
    Binney J., Tremaine S., 1987, Galactic Dynamics,
    Princeton University Press, Princeton 
\bibitem[Chandrasekhar(1943)]{1943ApJ....97..255C}
    Chandrasekhar S.\ 1943, \apj, 97, 255
\bibitem[Dekker(1974)]{1976PhR....24..315D}
    Dekker E., 1976, \prd, 24, 315
\bibitem[Fuchs(2004)]{2004A&A...419..941F}
    Fuchs, B.\ 2004, \aap, 419, 941
\bibitem[Fuchs \& Athanassoula(2005)]{2005AA...444..455F}
    Fuchs B., \& Athanassoula, E.\ 2005, \aap, 444, 455
\bibitem[Gradshteyn \& Rhyzik(2000)]{GR00}
    Gradshteyn, I.~S., \& Rhyzik, I.M.\ 2000, Table of Integrals, Series, and 
    Products, Academic Press, New York 
\bibitem[H\'enon(1973)]{H73}
    H\'enon, M.\ 1973, in: Dynamical structure and evolution of stellar systems,
    G. Contopoulos, M. H\'enon, D. Lynden-Bell (eds.),
    Lectures of the 3rd Advanced Course of the Swiss Society of Astronomy and
    Astrophysics, Geneva Observatory, Sauverny, p.~182
\bibitem[Hernquist(1990)]{1990ApJ...356..359H}
    Hernquist, L.\ 1990, \apj, 356, 359 
\bibitem[Kalnajs(1972)]{K72}
    Kalnajs, A.~J.\ 1972, in: Gravitational N--Body Problem, M. Lecar (ed.)
    Reidel Publ. Comp., Dordrecht, p.~13
\bibitem[Lynden-Bell \& Kalnajs(1972)]{1972MNRAS.157....1L}
    Lynden-Bell, D., \& Kalnajs, A.~J.\ 1972, \mnras, 157, 1 
\bibitem[Marochnik(1968)]{1968SvA....11..873M}
    Marochnik, L.~S.\ 1968, \sovast, 11, 873
\bibitem[Navarro, Frenk, \& White(1997)]{1997ApJ...490..493N}
    Navarro, J.~F., Frenk, C.~S., \& White, S.~D.~M.\ 1997, \apj, 490, 493
\bibitem[Stix(1992)]{ST92}
    Stix, T.~H.\ 1992, Waves in Plasmas, AIP, New York 
\bibitem[Tremaine \& Weinberg(1984)]{1984MNRAS.209..729T}
    Tremaine, S., \& Weinberg, M.~D.\ 1984, \mnras, 209, 729
\bibitem[White(1976)]{1976MNRAS.174...19W}
    White, S.~D.~M.\ 1976, \mnras, 174, 19
\end{thebibliography}
\end{document}